\documentclass[12pt]{iopart}

\usepackage{graphicx}
\usepackage[utf8x]{inputenc}
\usepackage{color}
\usepackage[dvipsnames]{xcolor}
\usepackage{subfig}
\usepackage{multirow}
\usepackage{slashed}
\usepackage{cancel}
\usepackage{stackrel}
\usepackage{amssymb}

\begin{document}

\title{Revisiting the rare $H\to q_iq_j$ decays in the Standard Model}

\author{J. I. Aranda$^{1}$, G. Gonz\'alez-Estrada$^{1}$, J. Monta\~no$^{1,2}$, F. Ram\'irez-Zavaleta$^{1}$, E. S. Tututi$^{1}$}

\address{
$^{1}$Facultad de Ciencias F\'isico Matem\'aticas, Universidad Michoacana de San Nicol\'as de Hidalgo,
Av. Francisco J. M\'ugica s/n, C.~P.~58060, Morelia, Michoac\'an, M\'exico.
\\
$^{2}$C\'atedras Conacyt, Av. Insurgentes Sur 1582, Col. Cr\'edito Constructor, Alc. Benito Ju\'arez, C.~P.~03940, Ciudad de M\'exico, M\'exico.
}

\ead{jmontano@conacyt.mx, feramirez@umich.mx}
\vspace{10pt}
\begin{indented}
\item[] Sep/15/2020
\end{indented}

\begin{abstract}
We revisit the rare decays of the Higgs boson to two different quarks in the Standard Model, which arise at the one-loop level. We perform Taylor series expansions to the complete form factors of the decay amplitudes, according to their different mass hierarchies, this allow us to take full advantage of the GIM mechanism to eliminate spurious contributions and retain those that truly contribute. We found that
Br$(H\to uc)$=$5.00\times10^{-20}$,
Br$(H\to ds)$=$1.19\times10^{-11}$,
Br$(H\to db)$=$5.16\times10^{-9}$ and
Br$(H\to sb)$=$1.15\times10^{-7}$.
Our predictions for the $H\to uc,ds$ decays disagree with previous results in the literature.
\end{abstract}

\section{Introduction}

The scalar boson observed at the LHC is compatible with that predicted by the Standard Model (SM)
\cite{Aad:2012tfa,Chatrchyan:2012xdj}, where by means of the Higgs Mechanism it is responsible for providing mass to the rest of the known elementary particles. The search for Higgs boson flavor changing neutral currents (FCNC) couplings, $\bar{q}_iq_jH$, is one of the topics of great interest in the particle physics community. This kind of interactions, also known as flavor violation in the Yukawa sector, is not present at the tree-level in the SM, nevertheless, the vertices arise as one-loop quantum fluctuations.
These interactions can be studied through the $H\to uc, ds, db, sb$ decays,
they involve the Glashow–Iliopoulos–Maiani (GIM) mechanism \cite{Glashow:1970gm}, which can severely suppress the FCNC.

As far as we know, in the context of the SM, in the Ref.~\cite{Bejar:2004rz} the branching ratio for the $H\to sb$ channel was estimated considering $m_H<2m_W$, being of order of $10^{-7}$. In addition, the Ref.~\cite{Benitez-Guzman:2015ana} is the only available study that has addressed all these four different Higgs decays taking into account the data at hand up to date. Nevertheless, the authors used \texttt{LoopTools}~\cite{Hahn:1998yk} to evaluate some complete Passarino-Veltman scalar functions (PaVe), which should not be done in this way because the PaVes contain spurious parts that can be removed by the GIM mechanism. This could strictly be achieved by splitting the PaVes, as a matter of fact the complete form factors, through Taylor expansions, which allows to keep the truly contributing parts in terms of the suppression fractions, or Inami-Lim terms \cite{Inami:1980fz}, $m_{q_k}^2/m_W^2\ll 1$, where $m_{q_k}$ is a virtual quark mass, except for the virtual top quark. Such issue has motivated us to recalculate the $H\to q_iq_j$ decays, which lead us to find discrepancies with two of the four results given in Ref.~\cite{Benitez-Guzman:2015ana}. It should be pointed out that the notation $H\to q_iq_j$ refers that the incoherent sum of the decay of the Higgs boson to the modes $q_i\bar{q_j}$ and $\bar{q_i}q_j$, which must be considered. On the other hand, by means of this approximation procedure we have successfully reproduced some SM processes where the GIM mechanism is involved, namely the top quark decays $t\to u_i\gamma, u_ig,~u_iZ,~u_iH$ \cite{AguilarSaavedra:2004wm}, and with the analogous leptonic GIM mechanism the decays $l_i\to l_j\gamma$ \cite{Cheng-Lee-book} and $\nu_i\to\nu_j\gamma$ \cite{Pal:1981rm}.

The paper is organized as follows. In Sec.~\ref{sec:Amplitude-loops} the $H\to uc$ and the $H\to ds,db,sb$ decays are calculated, the Taylor expansions are described and the GIM mechanism is applied in order to obtain the form factors free of spurious terms. In Sec.~\ref{SEC:predictions} we present the resulting expressions for the branching ratios and their numerical evaluations. The Sec.~\ref{SEC:Conclusions} is devoted to the conclusions. In the \ref{APPENDIX-Aux-func} some auxiliary functions are listed. In the \ref{APPENDIX-form-factors} the expanded form factors are evaluated in detail.

\section{The $H\to q_iq_j$ decays in the SM}
\label{sec:Amplitude-loops}

The Higgs decay into two distinct quarks $H\to q_iq_j$ consists of the incoherent sum of the two modes $H\to q_i\bar{q}_j$ and $H\to\bar{q}_iq_j$, whose width decays $\Gamma(H\to q_i\bar{q}_j)$ and $\Gamma(H\to\bar{q}_iq_j)$ are the same because of the invariance of the charge conjugation, parity and time reversal symmetry (CPT). Specifically, we will refer to them as $H\to q_iq_j\equiv H\to q_i\bar{q}_j+\bar{q}_iq_j$, which is an usual notation in the literature. In this sense, we establish the configuration  $H(p_3)\to q_i(p_1)\bar{q}_j(p_2)$, whose kinematics is $p_3=p_1+p_2$, $p_3^2=m_H^2$, $p_1^2=m_{q_i}^2$, $p_2^2=m_{q_j}^2$ and $p_1\cdot p_2=(m_H^2-m_{q_i}^2-m_{q_j}^2)/2$.

\subsection{The $H\to u_iu_j$ decay}
\label{SEC:Huiuj}

\begin{center}
\begin{figure}[ht]
\includegraphics[scale=1]{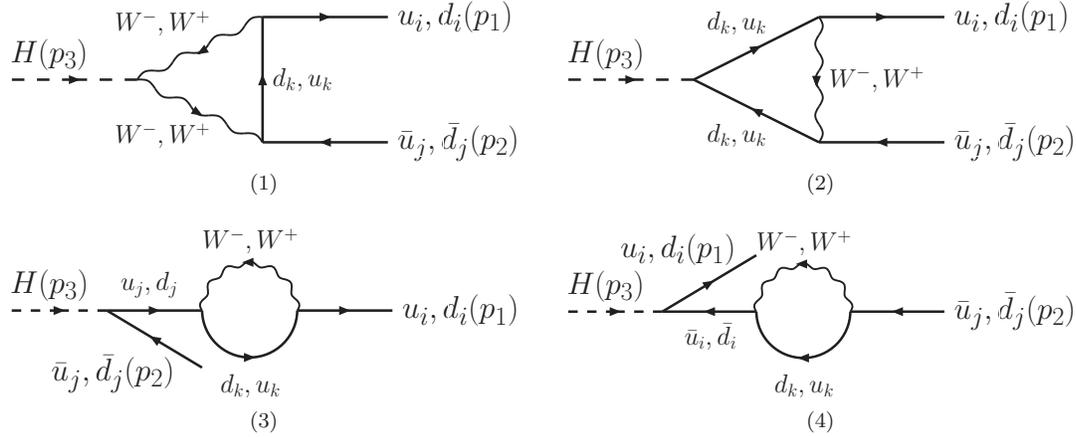}
\caption{The $H\to u_iu_j$ decay with $u_iu_j=u\bar c + \bar uc$ and $d_k=d,s,b$, and
$H\to d_id_j$ with $d_id_j=d\bar s+\bar ds,d\bar b+\bar db,s\bar b+ \bar sb$ and $u_k=u,c,t$.}
\label{FIGURE-loops1}
\end{figure}
\end{center}

The only possible Higgs decay into two distinct up-type quarks, $H\to u_iu_j$, corresponds to $u_iu_j=u\bar c+\bar u c$, comprised by the four Feynman diagrams depicted in the Fig.~\ref{FIGURE-loops1}, where inside the loops circulate the three down-type quarks $d_k=d_1,d_2,d_3=d,s,b$.
To calculate the corresponding loop integrals we have used the Passarino-Veltman tensor decomposition method \cite{Passarino:1978jh} through \texttt{FeynCalc}~\cite{Mertig:1990an,Shtabovenko:2016sxi} and \texttt{Package-X}~\cite{Patel:2016fam}. At this stage of calculation no approximations have been considered.

The resulting amplitude can be expressed as
\begin{equation}\label{Amplitude1}
\mathcal{M}= \bar{u}(p_1)\left(F_1+F_2\gamma^5\right)v(p_2)~,
\end{equation}
where the form factors $F_{1,2}$ have the generic structure
\begin{eqnarray}\label{FormFactorsCompleteuiuj}
F_a &=&\sum_{k=1}^3V_{u_id_k}V_{u_jd_k}^*
\left[
f_a^{A_1}A_0(1)+f_a^{A_2}A_0(2)+f_a^{B_1}B_0(1)+f_a^{B_2}B_0(2)
\right.
\nonumber\\
&&\left.
+f_a^{B_3}B_0(3)+f_a^{B_4}B_0(4)
+f_a^{C_1}C_0(1)+f_a^{C_2}C_0(2) \right],
\end{eqnarray}
with $a=1,2$, they depend on the PaVes
$A_0(1)$ $\equiv$ $A_0(m_{d_k}^2)$,
$A_0(2)$ $\equiv$ $A_0(m_W^2)$,
$B_0(1)$ $\equiv$ $B_0(m_{u_i}^2, m_{d_k}^2, m_W^2)$,
$B_0(2)$ $\equiv$ $B_0(m_{u_j}^2, m_{d_k}^2, m_W^2)$,
$B_0(3)$ $\equiv$ $B_0(m_H^2, m_{d_k}^2, m_{d_k}^2)$,
$B_0(4)$ $\equiv$ $B_0(m_H^2, m_W^2, m_W^2)$,
$C_0(1)$ $\equiv$ $C_0(m_{u_i}^2, m_{u_j}^2, m_H^2, m_W^2, m_{d_k}^2, m_W^2)$,
$C_0(2)$ $\equiv$ $C_0(m_{u_i}^2, m_{u_j}^2, m_H^2, m_{d_k}^2, m_W^2, m_{d_k}^2)$,
and the subform factors $f_a^{A1},...,f_a^{C2}$, which also depend on the masses of the particles.
At this point the amplitude (\ref{Amplitude1}) is ultraviolet divergent (UV), because there still remains the UV pole $1/\epsilon_\mathrm{UV}$ coming from the $A_0$ and $B_0$ PaVes.
In specific, the UV part has the form
\begin{eqnarray}\label{amplitudeUV}
\fl \mathcal{M}_\mathrm{UV}&=&-\!\sum_{k=1}^3V_{u_id_k}V_{u_jd_k}^*
~\frac{1}{\epsilon_\mathrm{UV}}~
\frac{ig^3m_H^2}{256\pi^2m_W^3}
\bar{u}(p_1)\!\left[(m_{u_i}+m_{u_j})-(m_{u_i}-m_{u_j})\gamma^5\right]\!v(p_2),
\end{eqnarray}
which does not depend on $m_{d_k}$, consequently, it is removed by the GIM mechanism $\sum_{k=1}^3V_{u_id_k}V_{u_jd_k}^*=0$.

In the following, in order to apply the GIM mechanism, the key Taylor expansion, on the form factors $F_a$, must be performed with respect to the virtual $q_k$ quark mass, provided $m_{q_k}/m_W\ll1$ be fulfilled. In addition, we have expanded with respect to the external $q_i$ and $q_j$ quark
masses due to $m_{q_i,q_j}/m_H\ll$ 1, which greatly simplify
the analytical results.

The $H\to u_iu_j$ decay, where $u_iu_j=u\bar c+\bar u c$, has virtual contribution of the three down-type quarks $d_k=d_1,d_2,d_3=d,s,b$. Considering the mass hierarchy $m_H$ $>$ $m_W$ $\gg$ $m_{u_i}$, $m_{u_j}$, $m_{d_k}$, we Taylor expand $F_a$ with respect to the smallest masses: the externals up to first order in $m_{u_i,u_j}/m_H\ll 1$, and the internals up to second order in $m_{d_k}/m_W\ll1$. After applying the GIM mechanism we get
\begin{eqnarray}\label{ff}
F_a &=& \sum_{k=1}^3V_{u_id_k}V_{u_jd_k}^*f_a(m_{d_k})~,
\end{eqnarray}
being
\begin{eqnarray}\label{f12-light-uiuj}
f_1(m_{d_k})&=&\frac{ig^3}{256\pi^2}\frac{m_{u_i}+m_{u_j}}{m_W}
\frac{\mathcal{F}_W}{1-r_W}\frac{m_{d_k}^2}{m_W^2}~,
\nonumber\\
f_2(m_{d_k})&=&\frac{-ig^3}{256\pi^2}\frac{m_{u_i}-m_{u_j}}{m_W}
\frac{\mathcal{F}_W}{1-r_W}\frac{m_{d_k}^2}{m_W^2}~,
\end{eqnarray}
where $\mathcal{F}_W$ and $r_W$ are defined in the \ref{APPENDIX-Aux-func}. Note that the leading Inami-Lim term $m_{d_k}^2/m_W^2\ll 1$ is quite small, which is the reason of the strongly suppressed value for $\mathrm{Br}(H\to uc)$$\sim$$10^{-20}$; it is explained in detail in Sec.~\ref{SEC:predictions} and in \ref{APPENDIX-form-factors} item i).

\subsection{The $H\to d_id_j$ decays}

The Higgs decay to two different down-type quarks, $H\to d_id_j$, with $d_id_j=d\bar s+\bar ds,d\bar b+\bar db,s\bar b+\bar sb$, has the virtual contribution of the three up-type quarks $u,c,t$, see Fig.~\ref{FIGURE-loops1}.
The amplitude for $H\to d_id_j$ is analogous to that for $H\to u_iu_j$ in Sec.~\ref{SEC:Huiuj}, when it is replaced $u_i\to d_i$, $u_j\to d_j$, $W^-\to W^+$, and $V_{u_id_k}V_{u_jd_k}^*\to V_{u_kd_i}^*V_{u_kd_j}$.
In order to apply the GIM mechanism in $F_a$, which satisfies $\sum_{k=1}^3V_{u_kd_i}^*V_{u_kd_j}=0$, to retain the dependent part on the $u_k$ mass, two different expansion schemes are required: one for the up and charm quarks, and the another one for the top quark. Then, to obtain the truly contributing parts as
\begin{eqnarray}\label{F12-didj-final}
F_a&=&\sum_{k=1}^3V_{u_kd_i}^*V_{u_kd_j}f_a(m_{u_k})~,
\end{eqnarray}
with $a=1,2$, the expansions should be performed as follows.

i) For the virtual contribution of the $u$ and $c$ quarks, where $m_H$ $>$ $m_W$ $\gg$ $m_{d_i}$, $m_{d_j}$, $m_u$, $m_c$, the expansion is similar to that implemented for $H\to u_iu_j$. The form factors are expanded with respect to the smallest masses, in specific, to the external ones because $m_{d_i,d_j}/m_H\ll 1$, and the internal ones since $m_{u_k}/m_W\ll 1$. Therefore, the $f_a$ can be written as
\begin{eqnarray}\label{f12-light-didj}
f_1(m_{u_k})&=&\frac{ig^3}{256\pi^2}\frac{m_{d_i}+m_{d_j}}{m_W}
\left(\frac{\mathcal{F}_W}{1-r_W}\frac{m_{u_k}^2}{m_W^2}+\mathcal{F}_W'\right),
\nonumber\\
f_2(m_{u_k})&=&\frac{-ig^3}{256\pi^2}\frac{m_{d_i}-m_{d_j}}{m_W}
\left(\frac{\mathcal{F}_W}{1-r_W}\frac{m_{u_k}^2}{m_W^2}+\mathcal{F}_W'\right),
\end{eqnarray}
for $u_k=u,c$; $\mathcal{F}_W'$ is given in the \ref{APPENDIX-Aux-func}.

ii) For the virtual contribution of the $t$ quark, the expansion can only be carried out with respect to the smallest masses $m_{d_i}$ and $m_{d_j}$, that is to say $m_{d_i,d_j}/m_H\ll 1$, then $f_a(m_t)$ results in
\begin{eqnarray}\label{f12-heavy-didj}
f_1(m_t)=\frac{ig^3}{256\pi^2}\frac{m_{d_i}+m_{d_j}}{m_W}~2\mathcal{F}_{Wt}~\frac{m_t^2}{m_W^2}~,
\nonumber\\
f_2(m_t)=\frac{-ig^3}{256\pi^2}\frac{m_{d_i}-m_{d_j}}{m_W}~2\mathcal{F}_{Wt}~\frac{m_t^2}{m_W^2}~,
\end{eqnarray}
being $\mathcal{F}_{Wt}$ defined in the~\ref{APPENDIX-form-factors}.

Summarizing, from Eqs.~(\ref{f12-light-didj}) and (\ref{f12-heavy-didj}) the final expressions for $F_a$ in Eq.~(\ref{F12-didj-final}) are
\begin{eqnarray}\label{F12-didj-final-final}
\fl F_1&=&
\frac{ig^3}{256\pi^2}\frac{m_{d_i}+m_{d_j}}{m_W}
\left[
\sum_{k=1}^2V_{u_kd_i}^*V_{u_kd_j}\left(\frac{\mathcal{F}_W}{1-r_W}\frac{m_{u_k}^2}{m_W^2}+\mathcal{F}_W'\right)
+V_{td_i}^*V_{td_j}2\mathcal{F}_{Wt}~\frac{m_{t}^2}{m_W^2}
\right],
\nonumber \\
\fl F_2 &=&
\frac{ig^3}{256\pi^2}\frac{m_{d_i}-m_{d_j}}{m_W}
\left[
\sum_{k=1}^2V_{u_kd_i}^*V_{u_kd_j}\left(\frac{\mathcal{F}_W}{1-r_W}\frac{m_{u_k}^2}{m_W^2}+\mathcal{F}_W'\right)
+V_{td_i}^*V_{td_j}2\mathcal{F}_{Wt}~\frac{m_{t}^2}{m_W^2}
\right].
\end{eqnarray}

\section{Predictions}
\label{SEC:predictions}

Because the $H\to q_iq_j\equiv H\to q_i\bar{q}_j+\bar{q}_iq_j$ process is due to the incoherent sum of the two final states $q_i\bar{q}_j$ and $\bar{q}_iq_j$, by CPT symmetry it results
\begin{eqnarray}\label{}
\Gamma(H\to q_iq_j)&=&\Gamma(H\to q_i\bar{q}_j)+\Gamma(H\to\bar{q}_iq_j),
\nonumber\\
&=&2~\Gamma(H\to q_i\bar{q}_j)~,
\end{eqnarray}
where
\begin{eqnarray}\label{width}
\Gamma(H\to q_i\bar{q}_j) &=& \frac{1}{16\pi m_H}
\sqrt{\left[1-\frac{(m_{q_i}+m_{q_j})^2}{m_H^2}\right]\left[1-\frac{(m_{q_i}-m_{q_j})^2}{m_H^2}\right]}~
|\mathcal{\overline{M}}|^2,
\nonumber\\
&\simeq&
\frac{1}{16\pi m_H}|\mathcal{\overline{M}}|^2,
\end{eqnarray}
with the mean squared amplitude
\begin{eqnarray}\label{square-amplitude}
|\mathcal{\overline{M}}|^2
&=& N_C\sum_\mathrm{spin}|\mathcal{M}|^2,
\nonumber\\
&\simeq&
2N_C m_H^2\left(|F_1|^2 +|F_2|^2\right).
\end{eqnarray}
The approximations in Eqs.~(\ref{width}) and (\ref{square-amplitude}) are possible because $m_H\gg m_{q_i},m_{q_j}$, therefore the branching ratio takes the form
\begin{eqnarray}\label{Br-generic}
{\mathrm{Br}(H\to q_iq_j)} &=& \frac{\Gamma(H\to q_iq_j)}{\Gamma_H}
\nonumber\\
&\simeq&
{\frac{N_Cm_H}{4\pi\Gamma_H}\left(|F_1|^2+|F_2|^2\right)},
\end{eqnarray}
where the  total decay width of the Higgs boson is $\Gamma_H=4.1\times10^{-3}$ GeV~\cite{Heinemeyer:2013tqa}.

For $H\to u_iu_j$, by considering Eqs.~(\ref{ff}) and (\ref{Br-generic}), the branching ratio can be expressed as
\begin{equation}\label{Br-H-uiuj}
\mathrm{Br}(H\to u_iu_j)=\frac{N_Cg^6m_H}{2^{17}\pi^5\Gamma_H}
\frac{m_{u_i}^2+m_{u_j}^2}{m_W^2}
\left|\frac{\mathcal{F}_W}{1-r_W}\sum_{k=1}^3V_{u_id_k}V_{u_jd_k}^*\frac{m_{d_k}^2}{m_W^2}\right|^2.
\end{equation}
where $u_iu_j=u\bar c+\bar uc$ and $d_k=d_1,d_2,d_3=d,s,b$.

For $H\to d_id_j$, by considering Eqs.~(\ref{F12-didj-final}) and (\ref{Br-generic}), the resulting branching ratio is
\begin{eqnarray}\label{Br-H-didj}
\mathrm{Br}(H\to d_id_j)&=&\frac{N_Cg^6m_H}{2^{17}\pi^5\Gamma_H}
\frac{m_{d_i}^2+m_{d_j}^2}{m_W^2}
\left|\sum_{k=1}^2V_{u_kd_i}^*V_{u_kd_j}\left(\frac{\mathcal{F}_W}{1-r_W}
\frac{m_{u_k}^2}{m_W^2}+\mathcal{F}_W'\right)
\right.
\nonumber\\
&&\left.
+V_{td_i}^*V_{td_j}2\mathcal{F}_{Wt}\frac{m_{t}^2}{m_W^2}\right|^2,
\end{eqnarray}
being $d_id_j=d\bar s+\bar ds,d\bar b+\bar d b,s\bar b+\bar sb$ and $u_k=u_1,u_2=u,c$.
\begin{table}[t!]
\begin{center}
\begin{tabular}{|c|c|}
\hline
$H\to q_{i}q_{j}$ & $\mathrm{Br}$        \\ \hline
$H\to uc$         & $5.00\times10^{-20}$ \\ \hline
$H\to ds$         & $1.19\times10^{-11}$ \\ \hline
$H\to db$         & $5.16\times10^{-9}$  \\ \hline
$H\to sb$         & $1.15\times10^{-7}$  \\ \hline			
\end{tabular}
\caption{Branching ratios of the $H \to q_{i}q_{j}\equiv H\to q_{i}\bar{q}_{j}+\bar{q}_{i}q_{j}$ decays.}
\label{Br}
\end{center}
\end{table}

Finally, our four predictions are summarized in the Table \ref{Br}. The data used in our numerical calculations come from the PDG 2020~\cite{PDG2020} and are displayed in the \ref{APPENDIX-Aux-func}. It should be noted that for the quark masses in final states we use their values at the Higgs boson mass scale, that is to say, we consider the running of the masses in the corresponding renormalization group equations (RGE), starting from the values of the quark masses given in the $\overline{\textrm{MS}}$ scheme in \cite{PDG2020} by using the \texttt{RunDec} package~\cite{Chetyrkin:2000yt,Herren:2017osy}. If we use the input data as in Ref.~\cite{Benitez-Guzman:2015ana}, our results for the $H\to d\bar b+\bar db,s\bar b+\bar sb$ decay channels agree with those reported in \cite{Benitez-Guzman:2015ana}, on the contrary, we have obtained different results for the $H\to u\bar c+\bar uc, d\bar s+\bar ds$ modes, for which they reported $\mathrm{Br}(H\to uc)\sim10^{-15}$ and $\mathrm{Br}(H\to ds)\sim10^{-8}$. It is worth appreciating why we disagree on two of the four predictions from Ref.~\cite{Benitez-Guzman:2015ana}. In fact, let us comment how they proceed: below their Eq.~(21) they recognize that the \texttt{LoopTools} software can not be used to evaluate the $B_0$ functions in a proper manner, because these could contain a $m_{q_k}$ independent part that should be removed by the GIM mechanism. Nevertheless, such statement must apply for any PaVe function dependent on the virtual quark mass $m_{q_k}$ if $m_{q_k}/m_W\ll 1$. Moreover, above their Eq.~(43) they state that their $C_0$'s were full evaluated with \texttt{LoopTools}, for all the cases, see their Table 2, therefore some of their evaluations are not appropriate.

However, in this work we employ, as pointed out above, the RGE running masses at the scale of the Higgs boson mass. Specifically, we found that Br$(H\to uc)=5.00\times10^{-20}$. In this case, the dominant part comes from the virtual $s$ and $b$ quarks, it is expected to be suppressed because of the Inami-Lim fraction $m_{d_k}^2/m_W^2\ll 1$ and by the $(m_u^2+m_c^2)/m_W^2\sim10^{-5}$ factor in Eq.~(\ref{Br-H-uiuj}); this branching ratio is the most affected by the GIM mechanism. For more details see \ref{APPENDIX-form-factors}.

In our prediction on Br$(H\to ds)=1.19\times10^{-11}$, the relevant contribution comes from the small virtual $u$ and $c$ quarks, since the virtual $t$ quark contribution is strongly suppressed by its own CKM matrix element $V_{td}^*V_{ts}\sim10^{-4}$, see Eq.~(\ref{Br-H-didj}) and \ref{APPENDIX-form-factors}; also the factor $(m_d^2+m_s^2)/m_W^2\sim10^{-7}$ helps to reduce its branching ratio. Here, the GIM mechanism removed the constant independent terms common to the three virtual quarks, nevertheless, it survives the constant $\mathcal{F}_W'$ in the $u$ and $c$ quark expansions. We recall that $m_t/m_W$ can not be Taylor expanded, then the corresponding form factors of the $t$ quark, analogous to those in Eq.~(\ref{FormFactorsCompleteuiuj}), can be directly evaluated with \texttt{LoopTools} as it was done in \cite{Benitez-Guzman:2015ana}, obviously this also applies for the $H\to d\bar b+\bar db,s\bar b+\bar sb$ decays.

For the $H\to d\bar b+\bar db,s\bar b+\bar sb$ processes, we obtain that Br$(H\to db)=5.16\times10^{-9}$ and Br$(H\to sb)=1.15\times10^{-7}$. These results are leading by the virtual top quark, that are not so suppressed by its CKM matrix elements (see \ref{APPENDIX-form-factors}). In this calculation, there is no expansion with respect to the virtual $m_t$, therefore the GIM mechanism can not be fully applied in this case, therefore, our evaluation coincides with that of Ref.~\cite{Benitez-Guzman:2015ana}.

\section{Conclusions}
\label{SEC:Conclusions}

We have presented compact analytical results for the rare $H\to u\bar c+\bar uc,d\bar s+\bar ds,d\bar b+\bar db,s\bar b+\bar sb$ decays in the SM at the one-loop level. Our predictions coincide with two of the four reported in \cite{Benitez-Guzman:2015ana}, we agree on the $H\to db,sb$ channels, by contrast, they reported $\mathrm{Br}(H\to uc)$$\sim$$10^{-15}$ and $\mathrm{Br}(H\to ds)$$\sim$$10^{-8}$, while we predict $\sim$$10^{-20}$ and $\sim$$10^{-11}$, respectively. The reason on why we disagree on two of the four decays from Ref.~\cite{Benitez-Guzman:2015ana} is because they did not properly separate the independent terms on the virtual $q_k$ masses and retain the dependent ones, which is mandatory to successfully apply the GIM mechanism, which results relevant for the $H\to u\bar c+\bar uc$ and $H\to d\bar s+\bar ds$ decays.

Our suppressed results are due to the fact that we have performed Taylor expansions to the full form factors of the decay amplitudes: one for $H\to u\bar c+\bar uc$ and another for $H\to d\bar s+\bar ds,d\bar b+\bar db,s\bar b+\bar sb$, which is justified by the different virtual quark mass hierarchies. The above mentioned allowed us to take full advantage of the GIM mechanism, that is to say, the spurious terms were removed.

Even though the experimental interest would be naturally directed to decays with the bottom quark in the final state, $H\to d\bar b+\bar db,s\bar b+\bar sb$, by completeness, it is worth knowing how and why the Higgs decaying into the lightest modes $u\bar c+\bar uc$ and $d\bar s+\bar ds$ is severely suppressed.

Furthermore, it is worth to highlight that in the decay modes $d\bar b+\bar db$ and $s\bar b+\bar sb$ the leading contribution comes from the virtual $t$ quark, hence these channels are relevant in the context of the $t$ quark physics. Since the top quark is the heaviest particle in the SM it is likely that it could have interactions with possible new heavy particles. So, if future detections of any of these FCNC rare decays $H\to d\bar b+\bar db,s\bar b+\bar sb$ differ from the SM prediction, it could be a symptom of new physics effects.

\section*{Acknowledgments}
This work has been partially supported by SNI-CONACYT and CIC-UMSNH. J. M. Thanks to professor J. J.  Toscano for useful discussions and C\'atedras Conacyt project 1753.

\appendix

\section{Auxiliary functions}
\label{APPENDIX-Aux-func}

The used input values from PDG 2020~\cite{PDG2020} are
$g=e/s_W$, $e=\sqrt{4\pi\alpha}$, $\alpha=\alpha(m_W)=1/128$, $s_W^2=s_W^2(m_Z)=0.23121$, the Higgs mass $m_H=125.1$ GeV, and the quark masses
$m_u = 0.00216$ GeV,
$m_d = 0.00467$ GeV,
$m_s = 0.093$ GeV,
$m_c = 1.27$ GeV,
$m_b = 4.18$ GeV,
$m_t = 172.76$ GeV.
Since these light quark masses are given in the $\overline{\textrm{MS}}$ scheme at low energy scales, we need their corresponding values at the energy scale of the Higgs mass. This can be achieved by using the \texttt{RunDec} package~\cite{Chetyrkin:2000yt,Herren:2017osy}, resulting in~\footnote{For this, we have used the following \texttt{RunDec} instructions: \texttt{AsRunDec} from the reference values $\alpha_s(m_Z)_\mathrm{PDG}$ and $m_Z$; the \texttt{mMS2mMS} command considering five active fermions and an accuracy of five loops.}:
$m_u = 0.001198$ GeV,
$m_d = 0.002591$ GeV,
$m_s = 0.05160$ GeV,
$m_c = 0.6033$ GeV,
$m_b = 2.799$ GeV. Also, the CKM matrix is
\[
\left(
\begin{array}{ccc}
|V_{ud}| & |V_{us}| & |V_{ub}| \\
|V_{cd}| & |V_{cs}| & |V_{cb}| \\
|V_{td}| & |V_{ts}| & |V_{tb}| \\
\end{array}
\right)
=
\left(
\begin{array}{ccc}
 0.9737 & 0.2245 & 0.00382 \\
 0.221 & 0.987 & 0.041 \\
 0.008 & 0.0388 & 1.013 \\
\end{array}
\right).
\]

In the following we present the constant functions $\mathcal{F}_W$, $\mathcal{F}_W'$ and $\mathcal{F}_{Wt}$, which depend on the masses $m_W$, $m_t$ and $m_H$.

The $\mathcal{F}_W$ appears in Eqs.~(\ref{f12-light-uiuj}), it is given by
\begin{eqnarray}\label{F-light}
\fl \mathcal{F}_W &\equiv& 2 (r_W-1) (\beta_W \ln2+\beta_W r_W \ln4-6 r_W+4)
+\pi ^2 (r_W-1)
\left(2 r_W^2+ r_W-1/3\right)
\nonumber\\
\fl && -2 i \pi  (r_W-1) (4 r_W-1)
+\left\{-2 (\beta_W+1)
+2 r_W [\beta_W (2 r_W-1)+3]
\right.
\nonumber\\
\fl && \left.
-2 i \pi  (r_W-1) r_W (4 r_W+1)\right\}\mathrm{l}_1
-(r_W-1) r_W (4 r_W+1)\mathrm{l}_1^2
\nonumber\\
\fl &&
-2 \beta_W (r_W-1) (2 r_W+1)\mathrm{l}_2
-2 (2 r_W-1)[-\beta_W+(\beta_W-3) r_W+1]\mathrm{l}_3
\nonumber\\
\fl && +2 (r_W-1) (2 r_W-1)
[(\beta_W-1) \mathrm{l}_4
-(\beta_W+1) \mathrm{l}_5]
\nonumber\\
\fl &&
+2 (2 r_W-1) [-\beta_W+(\beta_W+3) r_W-1]\mathrm{l}_6
-2 (r_W-1) r_W (4 r_W+1)\mathrm{L}_1
\nonumber\\
\fl &&
+2 (r_W-1)\left(2 r_W^2-r_W+1\right)(\mathrm{L}_2-\mathrm{L}_3+\mathrm{L}_4-\mathrm{L}_5+\mathrm{L}_6)
\nonumber\\
\fl &=& 3.943-2.565i,
\end{eqnarray}
where $r_W\equiv m_W^2/m_H^2=0.413$ and $\beta_W\equiv \sqrt{1-4r_W}=0.807i$, with the abbreviations of $\mathrm{l}_i$ and $\mathrm{L}_i$ listed below.

The $\mathcal{F}_W'$ appearing in Eqs.~(\ref{f12-light-didj}) is
\begin{eqnarray}\label{}
\mathcal{F}_W' &=& 4r_W(2r_W-1)\left(\mathrm{L}_2-\mathrm{L}_3+\mathrm{L}_4-\mathrm{L}_5+\mathrm{L}_6-\pi^2/6\right)
\nonumber\\
&=& 0.916~.
\end{eqnarray}

The $\mathcal{F}_{Wt}$ appears in Eq.~(\ref{f12-heavy-didj}), it is
\begin{eqnarray}\label{F-heavy-didj}
\mathcal{F}_{Wt}&\equiv&2 (r_W+r_t-1)+\mathrm{l}_1-\mathrm{l}_7+\beta_W  (2 r_W+1)\mathrm{l}_8-\beta_t  (4 r_W+2 r_t-1)\mathrm{l}_9
\nonumber\\
&&
-\frac{1}{r_t}\left[-4 r_W^3+2 r_W^2 (r_t+1)+r_W (2 r_t-1) r_t+r_t^2+r_t\right]m_H^2 C_0^t(1)
\nonumber\\
&&
+\left(4 r_W^2-2 r_W r_t+r_W-2 r_t^2\right) m_H^2 C_0^t(2)
\nonumber\\
&=&1.928~,
\end{eqnarray}
where $r_t\equiv m_t^2/m_H^2=1.907$, $\beta_t\equiv \sqrt{1-4r_t}=2.575i$ and $r_{Wt}\equiv r_W-r_t=-1.494$, and
\begin{eqnarray}\label{mH2C01qkheavy}
m_H^2 C_0^t(1)&=&m_H^2C_0(0, 0, m_H^2, m_W^2, m_t^2, m_W^2)
\nonumber\\
&=&\frac{1}{2}~\mathrm{l}_{10}^2+\mathrm{L}_{13}-\mathrm{L}_{14}+\mathrm{L}_{15}-\mathrm{L}_{16}+\mathrm{L}_{17}+\mathrm{L}_{18}
\nonumber\\
&=&-0.80288,
\end{eqnarray}
\begin{eqnarray}\label{mH2C02qkheavy}
m_H^2 C_0^t(2)&=&m_H^2C_0(0, 0, m_H^2, m_t^2, m_W^2, m_t^2)
\nonumber\\
&=&-\mathrm{L}_7+\mathrm{L}_8+\mathrm{L}_9-\mathrm{L}_{10}+\mathrm{L}_{11}-\mathrm{L}_{12}
\nonumber\\
&=& -0.40548.
\end{eqnarray}

The abbreviated logarithms $\mathrm{l}_i$ and dilogarithms $\mathrm{L}_i$ are
\begin{equation*}\label{}
\begin{array}{lll}
\mathrm{l}_1\equiv\ln r_W,
&
\mathrm{l}_2\equiv\ln(-\beta_W+2 r_W-1),
&
\mathrm{l}_3\equiv\ln\frac{\beta_W-1}{\beta_W-2 r_W+1}~,
\\
\\
\mathrm{l}_4\equiv\ln\frac{\beta_W+1}{\beta_W-2 r_W+1}~,
&
\mathrm{l}_5\equiv\ln\frac{\beta_W-1}{\beta_W+2 r_W-1}~,
&
\mathrm{l}_6\equiv\ln\frac{\beta_W+1}{\beta_W+2 r_W-1}~,
\\
\\
\mathrm{l}_7\equiv\ln r_t,
&
\mathrm{l}_8\equiv\ln\frac{2 r_W}{\beta_W+2 r_W-1}~,
&
\mathrm{l}_9\equiv\ln\frac{2r_t}{\beta_t+2 r_t-1}~,
\\
\\
\mathrm{l}_{10}\equiv\ln\frac{\beta_W+1}{\beta_W+2 r_{Wt}-1}~,
&
\mathrm{L}_1\equiv\mathrm{Li}_2(r_W+1),
&
\mathrm{L}_2\equiv\mathrm{Li}_2\frac{r_W-1}{r_W}~,
\\
\\
\mathrm{L}_3\equiv\mathrm{Li}_2\frac{2-2 r_W}{-2 r_W+\beta_W+1}~,
&
\mathrm{L}_4\equiv\mathrm{Li}_2\frac{-2 r_W}{-2 r_W+\beta_W+1}~,
&
\mathrm{L}_5\equiv\mathrm{Li}_2\frac{2 r_W-2}{2 r_W+\beta_W-1}~,
\\
\\
\mathrm{L}_6\equiv\mathrm{Li}_2\frac{2 r_W}{2r_W+\beta_W-1}~,
&
\mathrm{L}_7\equiv\mathrm{Li}_2\frac{r_{Wt}^2}{r_{Wt}^2+r_W}~,
&
\mathrm{L}_8\equiv\mathrm{Li}_2\frac{r_{Wt}^2+r_{Wt}}{r_{Wt}^2+r_W}~,
\\
\\
\mathrm{L}_9\equiv\mathrm{Li}_2\frac{-2 r_{Wt}}{-2 r_{Wt}+\beta_t-1}~,
&
\mathrm{L}_{10}\equiv\mathrm{Li}_2\frac{2 r_{Wt}+2}{2 r_{Wt}-\beta_t+1}~,
&
\mathrm{L}_{11}\equiv\mathrm{Li}_2\frac{2 r_{Wt}}{2 r_{Wt}+\beta_t+1}~,
\\
\\
\mathrm{L}_{12}\equiv\mathrm{Li}_2\frac{2 r_{Wt}+2}{2 r_{Wt}+\beta_t+1}~,
&
\mathrm{L}_{13}\equiv\mathrm{Li}_2\frac{r_{Wt}^2-r_{Wt} }{r_{Wt}^2+r_t}~,
&
\mathrm{L}_{14}\equiv\mathrm{Li}_2\frac{r_{Wt}^2}{r_{Wt}^2+r_t}~,
\\
\\
\mathrm{L}_{15}\equiv\mathrm{Li}_2\frac{2-2 r_{Wt}}{\beta_W+1}~,
&
\mathrm{L}_{16}\equiv\mathrm{Li}_2\frac{2-2 r_{Wt}}{-2 r_{Wt}+\beta_W+1}~,
&
\mathrm{L}_{17}\equiv\mathrm{Li}_2\frac{-2 r_{Wt}}{-2 r_{Wt}+\beta_W+1}~,
\\
\\
\mathrm{L}_{18}\equiv\mathrm{Li}_2\frac{2 r_{Wt}}{2 r_{Wt}+\beta_W-1}~.
&
&
\end{array}
\end{equation*}

\section{Evaluation of the form factors}
\label{APPENDIX-form-factors}

Here, we detail the evaluation of the form factors $|F_1|^2+|F_2|^2$ appearing in Eq.~(\ref{Br-generic}) of the generic branching ratio formula. It is sufficient to explicitly show the $F_1$ evaluation, since the only difference between $F_1$ and $F_2$ is a proportional factor, being $F_1\propto m_{q_i}+m_{q_j}$ and $F_2\propto -(m_{q_i}-m_{q_j})$, hence they are similar, $F_1\gtrsim F_2$; they can be found in the Eq.~(\ref{ff}) for $H\to uc$, and in Eq.~(\ref{F12-didj-final-final}) for $H\to ds,db,sb$. Thus, we show the explicit participation of each virtual quark contribution in the decay channels as follows.

i)~ For the $H\to uc$ decay, from Eq.~(\ref{ff}), the input data are:
$V_{ud}V_{cd}^*~f_1(m_d)=(2.15\times10^{-1})
(3.74+5.75i)\times10^{-15}$,
$V_{us}V_{cs}^*~f_1(m_s)=(2.22\times10^{-1})
(1.48+2.28i)\times10^{-12}$,
$V_{ub}V_{cb}^*~f_1(m_b)=(1.57\times10^{-4})
(4.36+6.71i)\times10^{-9}$. Thus,
\begin{eqnarray}
F_1 &=& V_{ud}V_{cd}^*~f_1(d)+V_{us}V_{cs}^*~f_1(s)+V_{ub}V_{cb}^*~f_1(b)
\nonumber\\
&=& (1.01+1.56i)\times10^{-12},
\end{eqnarray}
and the same pattern occurs for
\begin{eqnarray}
F_2 &=& (1.01+1.55i)\times10^{-12}.
\end{eqnarray}
Then, $|F_1|^2+|F_2|^2=6.87\times10^{-24}$, this in Eq.~(\ref{Br-generic}) yields $\mathrm{Br}(H\to uc)=5.00\times10^{-20}$.

\

ii)~ For $H\to ds$, from Eq.~(\ref{F12-didj-final}), by using the data:
$V_{ud}^*V_{us}~f_1(m_u)=(2.19\times10^{-1})(7.17\times10^{-17}+6.76\times10^{-8}i)$,
$V_{cd}^*V_{cs}~f_1(m_c)=(2.18\times10^{-1})(1.82\times10^{-11}+6.76\times10^{-8}i)$,
$V_{td}^*V_{ts}~f_1(m_t)=(3.10\times10^{-4})(1.32\times10^{-6}i)$,
it is obtained that
\begin{eqnarray}\label{F1-H-ds}
F_1 &=& V_{ud}^*V_{us}~f_1(m_u)+V_{cd}^*V_{cs}~f_1(m_c)+V_{td}^*V_{ts}~f_1(m_t)
\nonumber\\
&=& 3.96\times10^{-12}+2.99\times10^{-8}i,
\nonumber\\
F_2 &=& 3.58\times10^{-12}+2.71\times10^{-8}i.
\end{eqnarray}
The virtual $c$ quark provides the higher value. Hence, $|F_1|^2+|F_2|^2=1.63\times10^{-15}$ and $\mathrm{Br}(H\to ds)=1.19\times10^{-11}$.

\

iii)~ For $H\to db$, from Eq.~(\ref{F12-didj-final}), with the following values:
$V_{ud}^*V_{ub}~f_1(m_u)=(3.72\times10^{-3})(3.70\times10^{-15}+3.50\times10^{-6}i)$,
$V_{cd}^*V_{cb}~f_1(m_c)=(9.06\times10^{-3})(9.39\times10^{-10}+3.50\times10^{-6}i)$,
$V_{td}^*V_{tb}~f_1(m_t)=(8.10\times10^{-3})(6.80\times10^{-5}i)$, we get that
\begin{eqnarray}
F_1 &=& V_{ud}^*V_{ub}~f_1(m_u)+V_{cd}^*V_{cb}~f_1(m_c)+V_{td}^*V_{tb}~f_1(m_t)
\nonumber\\
&=& 8.51\times10^{-12}+5.96\times10^{-7}i,
\\
F_2 &=& 8.50\times10^{-12}+5.95\times10^{-7}i.
\end{eqnarray}
Here, the virtual $t$ quark provides the largest contribution. Therefore, $|F_1|^2+|F_2|^2=7.08\times10^{-13}$ and $\mathrm{Br}(H\to db)=5.16\times10^{-9}$.

iv)~ For $H\to sb$, from Eq.~(\ref{F12-didj-final}), by using the following data:
$V_{us}^*V_{ub}~f_1(m_u)=(8.58\times10^{-4})(3.77\times10^{-15}+3.56\times10^{-6}i)$,
$V_{cs}^*V_{cb}~f_1(m_c)=(4.05\times10^{-2})(9.56\times10^{-10}+3.56\times10^{-6}i)$,
$V_{ts}^*V_{tb}~f_1(m_t)=(3.93\times10^{-2})(6.92\times10^{-5} i)$, it is found that
\begin{eqnarray}\label{}
F_1 &=& V_{us}^*V_{ub}~f_1(m_u)+V_{cs}^*V_{cb}~f_1(m_c)+V_{ts}^*V_{tb}~f_1(m_t)
\nonumber\\
&=& 3.87\times10^{-11}+2.87\times10^{-6}i,
\nonumber\\
F_2 &=& 3.73\times10^{-11}+2.76\times10^{-6}i.
\end{eqnarray}
Again, the virtual $t$ quark yields the largest contribution. Thus, $|F_1|^2+|F_2|^2=1.58\times10^{-11}$ and $\mathrm{Br}(H\to sb)=1.15\times10^{-7}$.


\

\noindent{\bf References}

\

\end{document}